\documentclass[aps,prb,amsmath,twocolumn,amssymb,floatfix,showpacs,
superscriptaddress,footinbib,longbibliography]{revtex4-1}

\pdfoutput=1
\usepackage[dvips]{graphics}
\usepackage{bm}
\usepackage{epsfig}
\usepackage{enumerate}
\usepackage{subfigure}
\usepackage{color}
\usepackage{amsmath}
\usepackage{graphicx}
\usepackage{hyperref}
\hypersetup{
   pdfnewwindow=true,      % links in new window
   colorlinks=true,       % false: boxed links; true: colored links
   linkcolor=blue,          % color of internal links
   citecolor=magenta,        % color of links to bibliography
    filecolor=blue,      % color of file links
    urlcolor=blue        % color of external links
}

\newcommand{\bea}{\begin{eqnarray}}
\newcommand{\eea}{\end{eqnarray}}
\newcommand{\beq}{\begin{equation}}  
\newcommand{\eeq}{\end{equation}}

\begin{document} 
\title{Edge state reconstruction in proximitized graphene and its probe using RKKY interaction} 

\author{Seerat Khan}
\email{seerat.physics@gmail.com}
\affiliation{Jamia Millia Islamia, New Delhi-110025, India}
\author{SK Firoz Islam}
\email{s\_islam2@jmi.ac.in}
\affiliation{Jamia Millia Islamia, New Delhi-110025, India}

\begin{abstract}
Proximity induced spin-orbit coupling (SOC)  in graphene emerges when it is placed on a substrate, breaking sublattice symmetry and giving rise to staggered intrinsic SOC together with Rashba spin-orbit coupling. In zigzag graphene nanoribbons, this staggered SOC produces pseudohelical edge states exhibiting sublattice-locked rather than spin-locked helicity. We investigate the effect of a finite on-site potential, restricted to the edges of the ribbon, on these edge states. We find the inter-valley edge states undergo reconstruction and become purely intra-valley, exhibiting
both spin and valley polarization. The charge conductance calculations fail to capture this reconstruction, since the number of transmitted channels remains unchanged. We show that the Ruderman-Kittel-Kasuya-Yosida (RKKY) magnetic exchange interaction, between two magnetic impurities situated on the ribbon edge can be used to probe the shifted band dispersion, for both the uniform and staggered spin-orbit coupling. In the presence of strong intrinsic and Rashba SOC, the RKKY interaction becomes fully anisotropic. Both sublattice staggering and edge potential-induced reconstruction lead to distinct features in the RKKY interaction. We show that the RKKY interaction terms acquire faster oscillations as a function of the spatial separation between two magnetic impurities, in response to the edge state reconstruction, establishing it as an effective probe of the edge states. Additionally, the significant mismatch in the amplitudes of the DM term corresponding to the uniform and staggered SOC, can also act as a probe of the type of SOC.

\end{abstract}

\maketitle
%-------------------------
\section{Introduction}
%-------------------------
Spin orbit interactions (SOI) are essential for realising many unique phenomena in solid state systems, with applications ranging from data storage to information processing\cite{Fabian2004}. The discovery and isolation of graphene opened a new avenue for spintronics based devices, owing to the excellent charge and spin transport  properties of Dirac electrons \cite{castro2009}. However, the intrinsic spin-orbit coupling (SOC) in graphene is too weak to offer any meaningful control over its spin degree of freedom\cite{Gmitra,Sichau}. Various methods have been suggested to enhance the strength of SOC\cite{Han2014,Castro}; one of the most effective routes is to place graphene on a substrate leading to proximity induced effects\cite{Zollner,Sierra,Zutic}. Some of the substrate properties are transferred to graphene and the strength of SOC increases dramatically.  Experiments with transition metal dichalcogenides (TMDs) like WSe$_2$ and MoS$_2$ have shown to increase the SOC of graphene upto a few meVs\cite{Ghiasi,Wang,Omar}. 
Combining graphene  with a 2D ferromagnetic insulator Cr$_2$Ge$_2$Te$_6$ , induces an out-of-plane magnetic exchange interaction in graphene, suggesting a possible non-trivial spin texture\cite{Karpiak}.  Proximity effects induced by substrates like TMDs also change the functional form of SOC by breaking the sublattice symmetry,  due to which A and B sublattices experience opposite SOC coupling (valley Zeeman splitting)\cite{benitez,Cummings,Ghiasi}. Additionally, an orbital gap opens due to staggered on-site potential, and induction of Rashba SOC breaks horizontal reflection symmetry. When staggered SOC is incorporated in the modified Haldane model\cite{Haldane}, pseudo-helical edge states are formed \cite{Frank}. These new pseudospin-spin locked states are different from the helical edge modes of quantum spin Hall state (QSHS) in the Kane and Mele model\cite{Kane}, in the sense that now same spin states travel in same  direction on opposite edges leading to a net spin current. These proximity effects can also be controlled electrically via gate voltage, to provide increased functionality. Recent studies suggest that reversing gate voltages can electrically switch either spin and/or valley polarization in chirally stacked graphene encapsulated between TMD layers\cite{Taige}. The on-site edge potential confined to the geometrical edge acts differently from bulk gate voltage. In zigzag nanoribbons particularly, an edge localized potential can couple the spin and valley degrees of freedom, giving rise to distinct spin–valley polarized phases\cite{Lu}.

In this work, we consider two cases individually, the zigzag graphene nanoribbon with uniform SOC and with staggered SOC. In both cases, we also keep the Rashba SOC term. We first briefly revisit the edge states that have been revealed by Fabian group in Ref.~[\onlinecite{Frank}]. Subsequently, we apply an on-site edge potential to reconstruct the edge states. 
In the uniform SOC case, after applying edge potential, the inter-valley edge state crossings of the QSHI shift towards the valley gap and we are left with purely intra-valley edge states. The system acquires valley polarization in addition to spin polarization. In the staggered case, where the pseudohelical edge states are already spin-valley locked, applying edge potential similarly reconstructs them into intra-valley edge states. Next we try to extract the signatures of edge modes in presence and absence of edge potential, corresponding to uniform and staggered SOC cases, by computing longitudinal conductance.  We note that the charge conductance fails to capture the signature of these reconstructed edge states in either case, since the number of low energy transmitted channels remains unchanged. Therefore we require a probe directly sensitive to the spin and valley character of the edge states. We show that the RKKY exchange interaction could be a better probing tool for such edge states. The RKKY interactions have already been know as a sensitive and powerful probe for the edge states and topological phases in related systems \cite{Zare,Islam}, including the perturbed Kane mele model \cite{Mohsen,sayyid} and helical Majorana edge modes in topological superconductors \cite{Katharina}. The RKKY interactions become anisotropic and electrically tunable when the mediating states carry spin-orbit coupling, acquiring a nontrivial tensor structure dictated by the residual spin symmetries of the system \cite{Lee}. We compute the RKKY interactions between two magnetic impurities, placed on the ribbon edge, as a function of the distance between them to extract the signature of edge potential reconstruction in the edge modes. We also consider the cases when the two impurities are situated on same (AA) and different (AB) sublattice sites of the zigzag ribbon edge. Our results show that the application of an edge potential enhances the oscillation frequency of the RKKY interaction terms. In the presence of staggered SOC, the Dzyaloshinskii–Moriya (DM) interaction exhibits higher amplitude, accompanied by the emergence of nonzero in-plane components. Additionally, one of the symmetric exchange terms show a beating pattern when both staggered SOC and edge potential are present.

This study is organised as follows. In Sec.~\ref{sec2} we describe our tight binding model hamiltonian with the inclusion of edge potential. In Sec.~\ref{sec3}, we analyse the band spectrum for uniform and staggered SOC and the effect of edge potential on the ribbon edges. In Sec.~\ref{probeA} we analyse the charge conductance calculated using Green’s function formalism. In Sec.~\ref{probeB} we build the theoretical framework for RKKY exchange interaction and present the numerical results for various terms of RKKY interaction with and without edge potential, for both uniform and staggered SOC regimes. Finally, we summarise and conclude our analysis in Sec.~\ref{conc}.

%%%%%%%%%%%%%%%%%%%%%%%% Ribbon type %%%%%%%%%%%%%%%%%%%
\begin{figure}[!htb]
\centering
\includegraphics[width=1.0\linewidth]{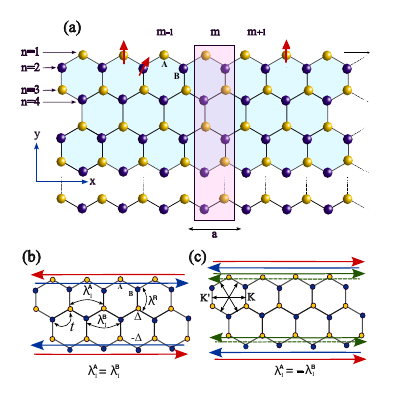}
\caption{(Color online) Fig. (a) illustrates the lattice configuration of zigzag nanoribbon. The pink rectangular box denotes a unit cell and $m$, $n$ represent the $x$ and $y$ coordinates of the lattice points. The ribbon dimensions are defined by $L$, representing the number of unit cells along its length and the width  $N_y$, denoting the number of atoms per unit cell. The magnetic impurities are depicted by the vertical red arrow signs. Fig. (b) and (c) show a schematic illustration of proximity-induced properties in graphene. The black labeled elements refer to bulk graphene properties common to both cases. Sublattice A is shown as yellow circles and sublattice B as blue circles. Fig. (b) illustrates the tight-binding hopping parameters used in our model. The long arrows indicate the helical edge states along with their propagation directions for the uniform case ($\lambda_I^A=\lambda_I^B$), with red and blue symbols denoting spin-up and spin-down character, respectively. Fig. (c) shows the same lattice with the reciprocal-space $K$ and $K'$ directions overlaid, now for the staggered intrinsic SOC case ($\lambda_I^A=-\lambda_I^B$). Here, the solid and dashed green arrows mark the valley edge states belonging to the $\kappa=1$ and $\kappa=-1$ valleys, respectively, while the red and blue arrows indicate the pseudohelical states, which carry a net spin current along the ribbon.}

\label{ribbon}
\end{figure}
%%%%%%%%%%%%%%%%%%%%%%%%%%%%%%%%%%%%%%%%%%%%%%%%%%%%%%%%

%----------------------------------------------------------------------
\section{Tight binding Model Hamiltonian}\label{sec2}
%----------------------------------------------------------------------
The tight-binding Hamiltonian for graphene  placed on a substrate, including sublattice asymmetry and a staggered potential, is given by~\cite{Frank}:

\begin{equation}
\begin{aligned}
\mathcal{H}_0 ={}&
\sum_{\langle i,j\rangle;s} t c_{is}^{\dagger}c_{js}
+ \sum_{i,s} \Delta \xi_i c_{is}^{\dagger}c_{is} \\
&+ \frac{2\lambda_R}{3}i\sum_{\langle i,j\rangle,ss'} c_{is}^{\dagger}c_{js'}
\left[(\hat{\mathbf{s}}\times\mathbf{d}_{ij})_z\right]_{ss'} \\
&+ \frac{i}{3\sqrt{3}}\sum_{\langle\langle i,j\rangle\rangle,ss'} \lambda_I^i \nu_{ij} c_{is}^{\dagger}c_{js'}
\left[\hat{s}_z\right]_{ss'} .
\end{aligned}
\label{eq:Hamiltonian}
\end{equation}
where operator $c_{is}^{\dagger}(c_{is})$ creates (annihilates) an electron with spin $s$ at site $i$, belonging to sublattice A or B. The first term describes nearest-neighbor hopping (sum over $\langle i,j\rangle$) with amplitude $t$, which conserves the spin index $s$. The second term represents the staggered sublattice potential, where $\Delta$ is its magnitude and $\xi_i=+1$ and $-1$, for sites belonging to sublattices A and B, respectively.

The third term corresponds to Rashba SOC with coupling strength $\lambda_R$. This term breaks horizontal reflection symmetry and mixes states with opposite spins and sublattices. It depends on the direction of the nearest-neighbor bond through the unit vector $\mathbf{d}_{ij}$, which points from site $j$ to site $i$. The vector $\hat{\mathbf{s}}$ denotes the spin Pauli matrices.

The last term represents intrinsic spin-orbit coupling, which involves next-nearest-neighbor hopping (sum over $\langle\langle i,j\rangle\rangle$) and couples same spins. The sign factor $\nu_{ij}=+1(-1)$ corresponds to counterclockwise (clockwise) hopping paths along a hexagonal ring from site $j$ to site $i$, respectively. The parameter $\lambda_I^i$ allows the intrinsic spin-orbit coupling to differ between the two sublattices, with $i$ denoting either sublattice A or B.
%--------------------------------------
\subsection{Inclusion of on-site edge potential}
%--------------------------------------

The band structure can be modified by application of suitable on-site edge potential on the opposite zigzag edges \cite{Chiu,Apel}. In order to tune the full band dispersion, we tune the top and bottom edges as $V_t=+V$ and $V_b=-V$. The total Hamiltonian of the system, including the effect of the on-site edge potential can be written as $\mathcal{H}=\mathcal{H}_0+\mathcal{H}_{edge}$ with
\begin{equation}
\mathcal{H}_{edge}=\sum_{i\in \text{top},s}V_t c_{is}^{\dagger}c_{is}+ \sum_{i\in \text{bottom},s}V_b c_{is}^{\dagger}c_{is}.
\tag{2}
\end{equation}
In the present case, we will keep $V_t=-V_b$.
%--------------------------------------
\section{Band dispersion}\label{sec3}
%---------------------------------------
\begin{figure}[htb]
    \centering
  
     % Second row
    \includegraphics[width=0.435\linewidth]{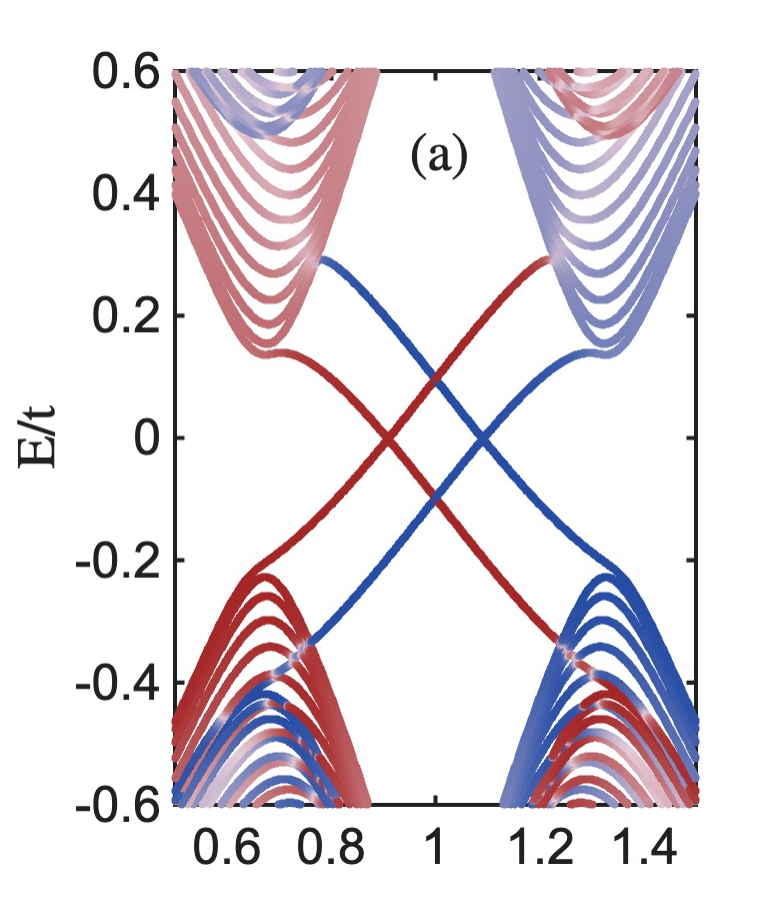}
    \includegraphics[width=0.41\linewidth]{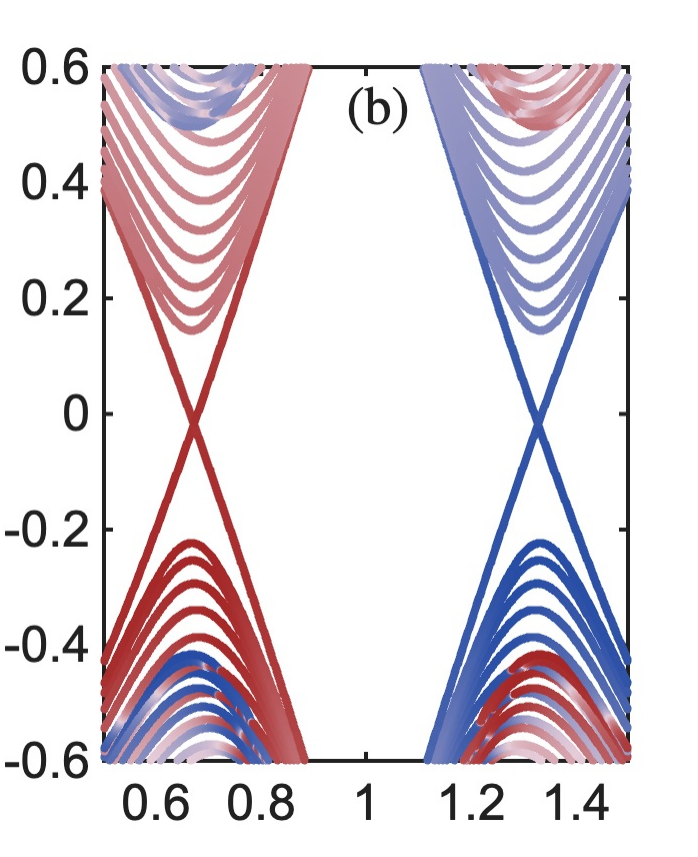}
    
    \includegraphics[width=0.45\linewidth]{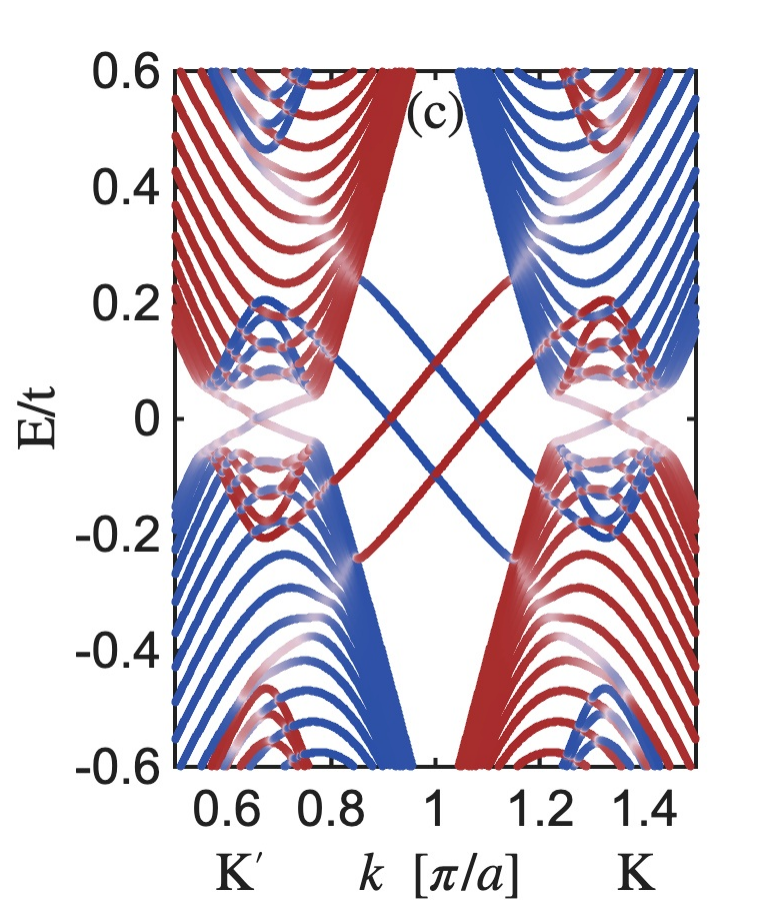}
    \includegraphics[width=0.41\linewidth]{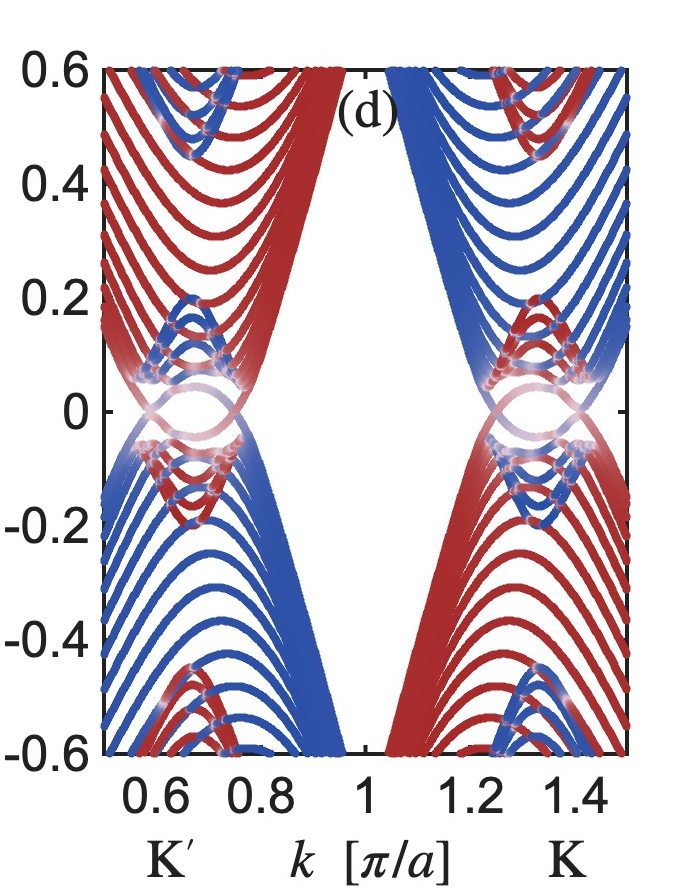}
\caption{(Color online) Energy spectrum of a zigzag ribbon of width $N_y=100$. The spin expectation values are depicted as red for spin-up, and blue for spin-down. (a), (b) show the uniform case, $\lambda_I^A = \lambda_I^B$ and (c),(d) show the staggered case, $\lambda_I^A = -\lambda_I^B$. The edge potential is fixed at $V=0$ in (a) and (c) and $V=t$ in (b) and (d).}
    \label{band}
\end{figure}
In this section we first briefly discuss the band structure of the proximitized graphene nanoribbon. The lattice geometry of the zigzag nanoribbon is illustrated in Fig.~\ref{ribbon}. The nanoribbon can be viewed as a periodic array of unit cells, represented by the pink shaded rectangular box (finite along $y$ and infinite along the $x$-direction). The width of the zigzag ribbon is determined by the number of atoms $N_y$ per unit cell. The energy eigenvalues can be obtained using an effective difference equation, similar to the case of an infinite one-dimensional chain having the following form
\begin{equation}
(EI-\mathcal{E})\psi_m=\mathcal{T}\psi_{m+1}+\mathcal{T}^{\dagger}\psi_{m-1}
\tag{3}
\label{eq3}
\end{equation}
where,
\begin{equation}
\psi_m=
\begin{bmatrix}
\psi_{m,1}\\
\psi_{m,2}\\
\vdots\\
\vdots\\
\psi_{m,N_y}
\end{bmatrix}
\tag{4}
\end{equation}
Here, $\mathcal{E}$ is the on-site energy matrix, $\mathcal{T}$ is nearest-neighbor hopping matrix between adjacent unit cells and $I$ is the identity matrix of dimension $2N_y\times2N_y$. Since the zigzag chain is translationally invariant along the $x$-direction, the corresponding momentum $k$ is conserved and serves as a good quantum number. The periodic nature of the ribbon along the $x$-direction allows us to apply the Bloch's theorem which yields the total Hamiltonian as
\begin{equation}
(EI-\mathcal{E})=\mathcal{T}e^{ika}+\mathcal{T}^{\dagger}e^{-ika}
\tag{5}
\end{equation}
where $a$ is the lattice constant. We numerically solve the above equation to obtain the energy dispersion of the nanoribbon. We use the numerical parameters as: $t=1$, $\Delta=0.1t$, $\lambda_R=0.075t$, and $\lvert\lambda_I^A\rvert,\lvert\lambda_I^B\rvert=\sqrt{27}\times 0.06t$, as used by Ref.~[\onlinecite{Frank}].

\begin{figure*}[tb]
    \centering

    % First row
    \hspace{3mm}
     \includegraphics[width=0.62\columnwidth]{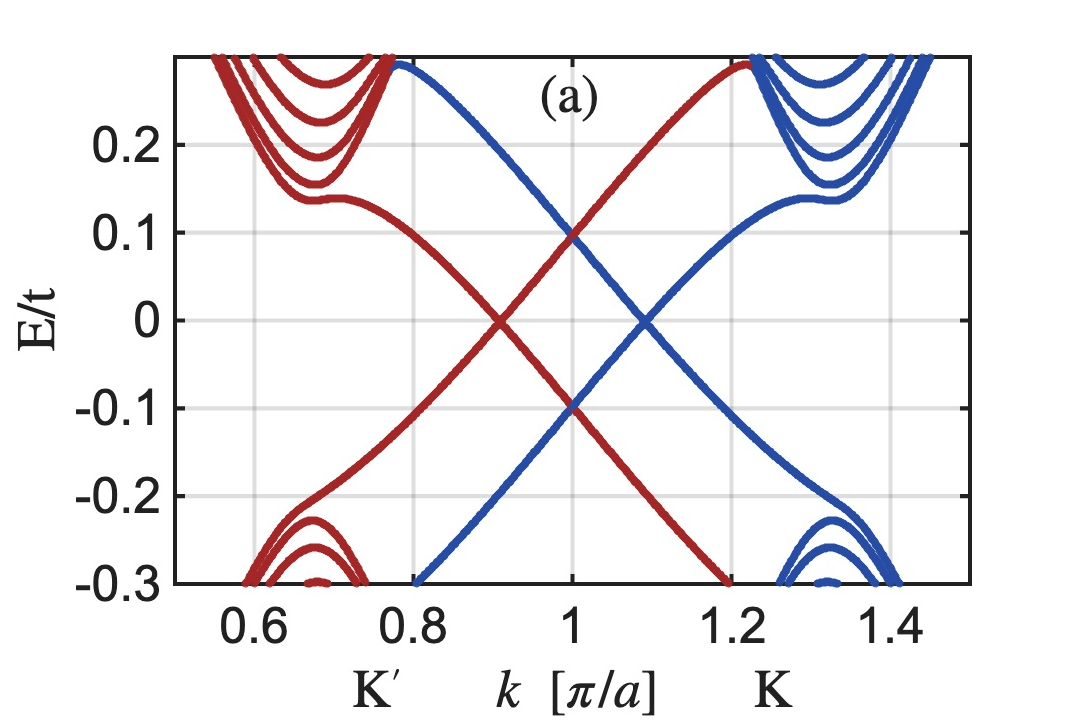}
    \hspace{-0.6cm}%
      \raisebox{0.8mm}{
    \includegraphics[width=0.30\columnwidth]{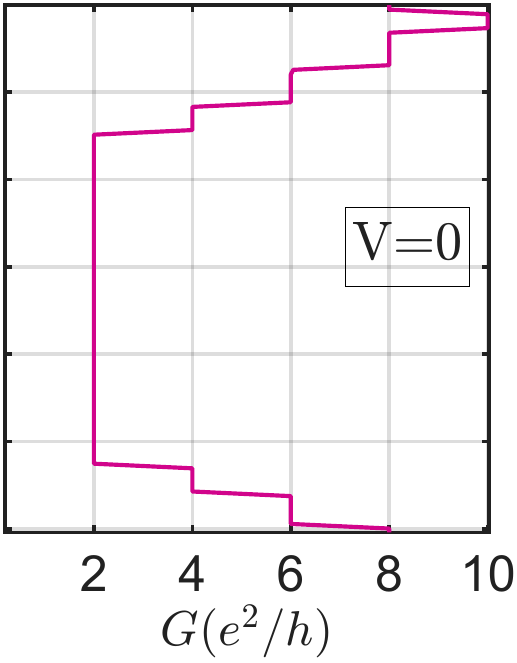}}
    \hspace{0.7cm}
    \includegraphics[width=0.63\columnwidth]{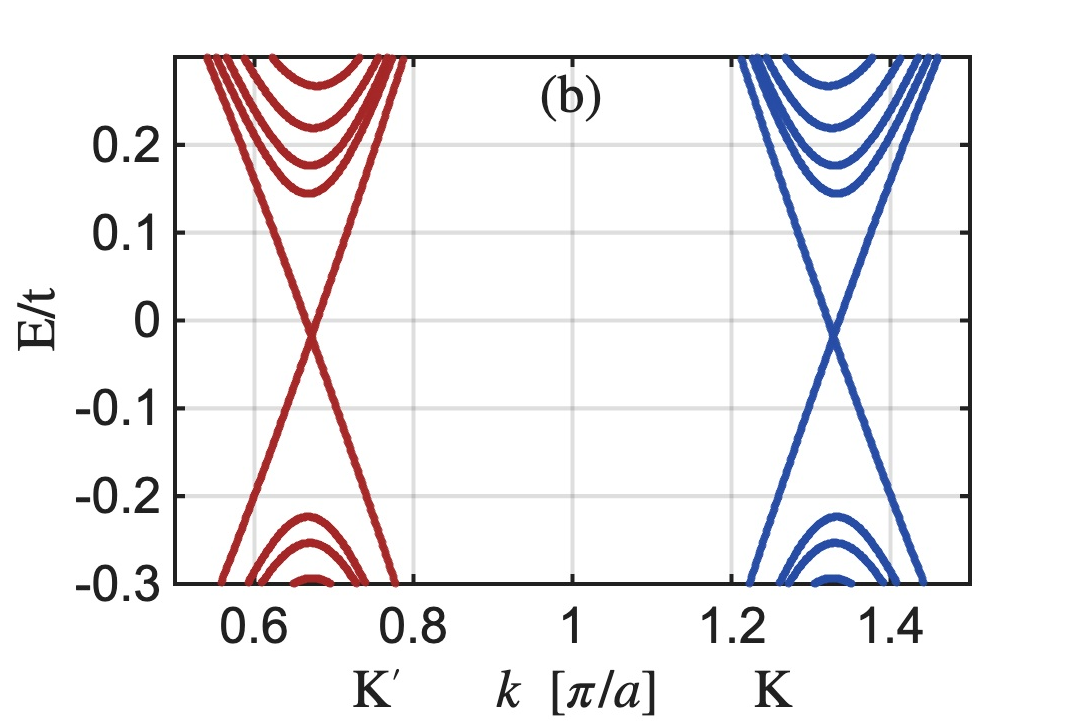}
     \hspace{-0.6cm}%
      \raisebox{0.8mm}{
    \includegraphics[width=0.30\columnwidth]{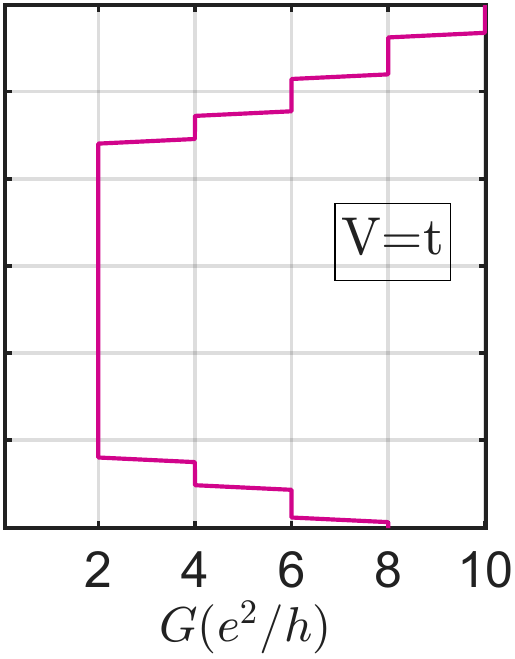}}
     %  \vspace{0.3cm}

   %  % Second row
   \hspace{2mm}
    \includegraphics[width=0.64\columnwidth]{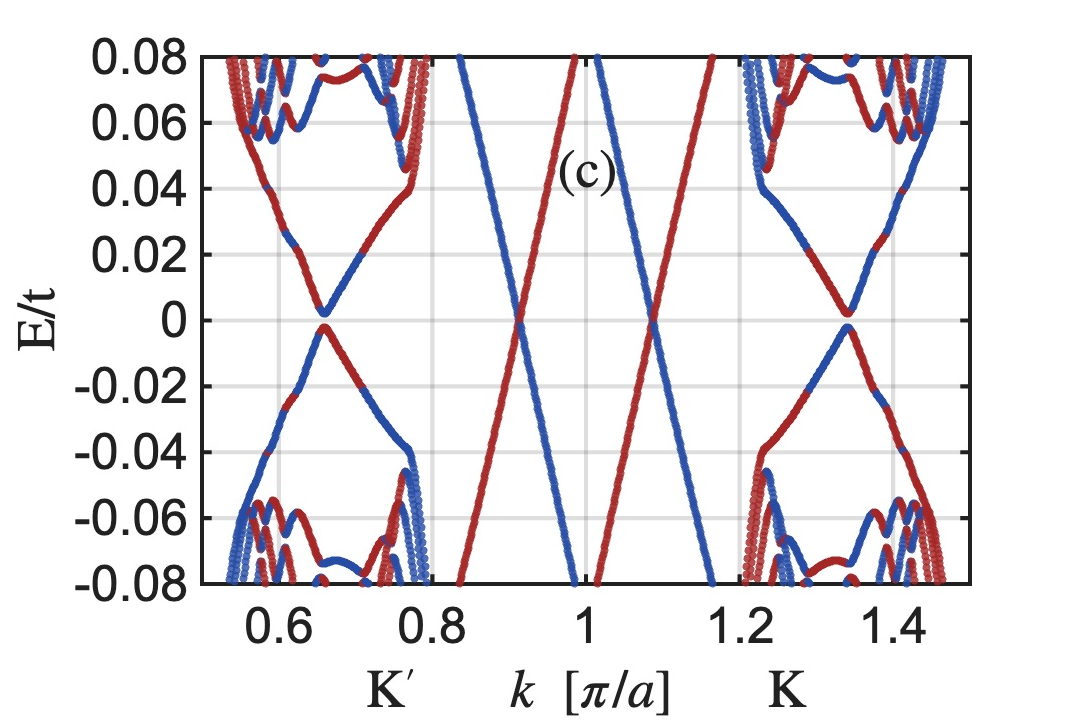}
    \hspace{-0.6cm}%
      \raisebox{0.8mm}{
    \includegraphics[width=0.31\columnwidth]{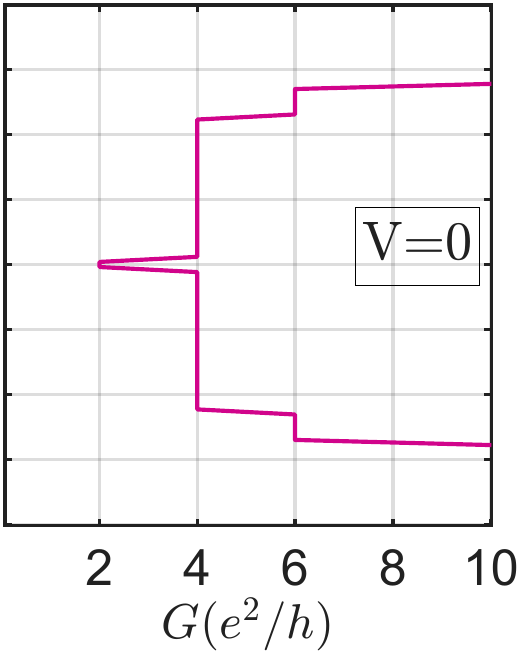}}
    \hspace{0.4cm}
    \includegraphics[width=0.64\columnwidth]{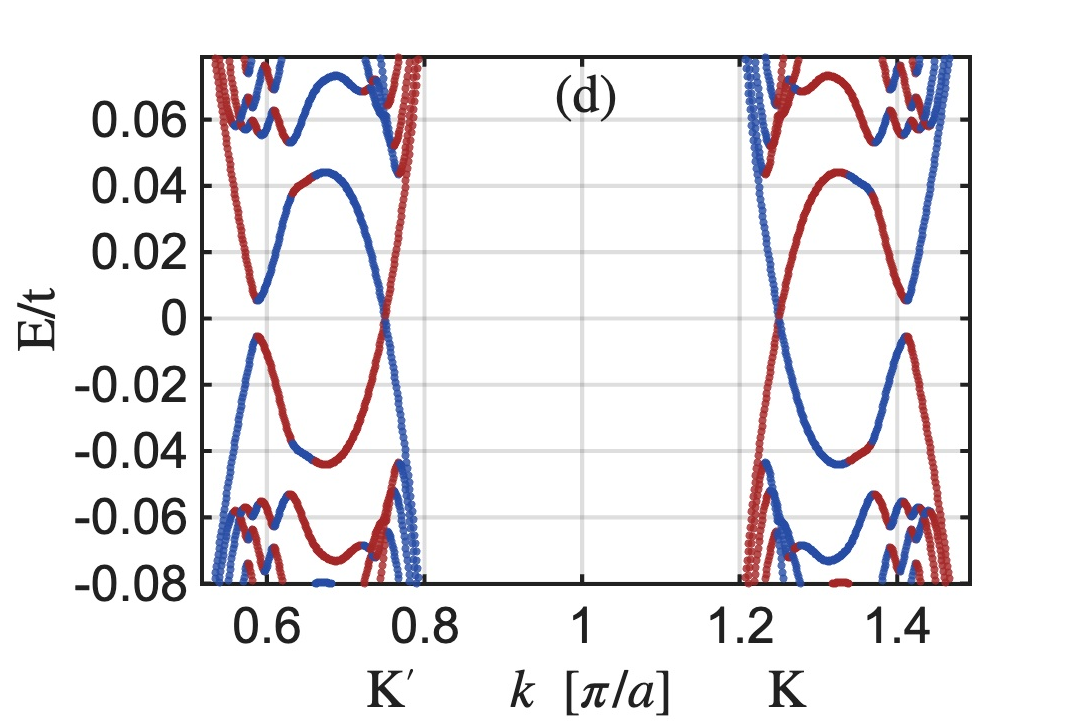}
     \hspace{-0.6cm}%
      \raisebox{0.8mm}{
    \includegraphics[width=0.31\columnwidth]{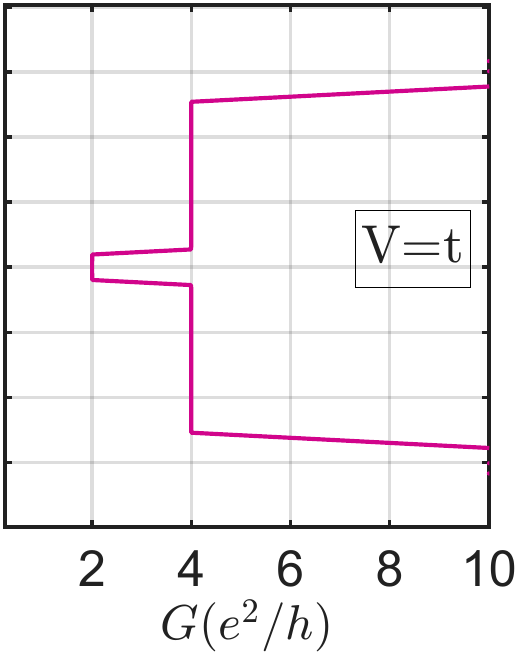}}
 \caption{(Color online) Low energy spectrum and charge conductance of incoming electrons for a zigzag ribbon of width $N_y=100$. Red and blue denote spin-up and spin-down characters, respectively. (a),(b) show the uniform case, $\lambda_I^A = \lambda_I^B$ and (c),(d) show the staggered case, $\lambda_I^A = -\lambda_I^B$. The edge potential is fixed at $V=0$ in (a) and (c) and $V=t$ in (b) and (d).}
    \label{cond}
\end{figure*}

In Fig.~\ref{band}, we show the energy band dispersion of a zigzag ribbon of width $N_y = 100$ for two values of edge potentials, $V=0$ in (a) and (c) and $V=t$ in (b) and (d). In the uniform case Fig.~\ref{band}(a), the two spin sectors combine additively and the band structure shows two pairs of helical edge states, this is the  known quantum spin Hall state (QSHS) \cite{Kane} [see Fig.~\ref{ribbon} (b)]. Inclusion of Rashba SOC causes mixing of spins in the bulk bands which is more dominant in the conduction bands. 

In the staggered case, Fig.~\ref{band}(c), in contrast to the QSHS, where same-spin states on opposite edges counter-propagate, here they move in the same direction, producing a net spin current as shown in Fig.~\ref{ribbon}(c). These states are termed as pseudohelical edge states \cite{Frank}. Inclusion of Rashba SOC opens a bulk gap in the valleys because the valence and conduction bands now have different spin expectation values. This is the mass inversion regime \cite{Alsharari}, with sections of the former valence bands now sitting higher in energy than sections of the former conduction bands. Two new edge states with quenched spins appear inside this Rashba gap for each valley depicted in Fig.~\ref{ribbon}(c), together with the pseudohelical pair this is defined as the quantum valley spin Hall state (QVSHS) \cite{Frank}. Additionally, a small gap exists at the valley states due to spectral proximity to the bulk states but for wide enough ribbons [see Fig.~\ref{cond} (c) and (d)], it should not affect the edge physics \cite{Frank}.

\subsection{Effect of edge potential}
Let us now turn on the on-site edge potential. The band structure for uniform and staggered cases is presented in Fig.~\ref{band} (b) and (d), respectively. It shows that the edge states do not connect the two valleys anymore, rather these are shifted towards the valley gap by some finite amount proportional to the edge potential, until they sit directly inside the valley gap for $V=t$. For the uniform case Fig.~\ref{band}(b), the inter-valley edge mode crossings now become strictly intra-valley and lead to valley-polarised edge modes. For the staggered case Fig.~\ref{band}(d), these intravalley crossings coexist with the additional valley states mentioned earlier. For narrow ribbons, the pre-existing valley states acquire a relatively higher band gap in the presence of edge potential, which increases gradually with decreasing $N_y$. This originates due to the combined effect of edge potential and the coupling between upper and lower edges via tunneling\cite{Lu}. However, the additional valley states due to shifted inter-valley edge states, remain gapless. 

\section{Probing edge state reconstruction}\label{probe}
In this section, we discuss two possible ways of extracting the signatures of edge state reconstruction. First, we consider the longitudinal conductance, then we investigate the RKKY exchange interaction between two magnetic impurities situated at the ribbon edge. Let us first evaluate the conductance.

\subsection{Conductance probe}\label{probeA}
 We compute the two-terminal conductance using the standard recursive Green's function method \cite{Shakouri,Dutta}. The conductance follows from the Landauer formula,
\begin{equation}
G = \frac{e^2}{h} \mathrm{Tr}\left[ \Gamma_L(E) G_D^{\dagger} \Gamma_R(E) G_D(E) \right] ,
\label{eq:conductance}
\tag{6}
\end{equation}
where  the broadening matrices including the coupling
between transport channel with left/right leads can be
written as
\begin{equation}
\Gamma_{L/R}(E) = i\left[ \Sigma_{L/R}(E) - \Sigma_{L/R}^{\dagger}(E) \right] ,
\tag{7}
\end{equation}
with $\Sigma_{L/R}(E)$ the corresponding lead self-energy. The retarded Green's function of the central (scattering) region is then
\begin{equation}
G_D = \left[ E - \mathcal{E} - \Sigma_L - \Sigma_R \right]^{-1} ,
\label{eq:greenD}
\tag{8}
\end{equation}
where $\mathcal{E}$ is the on-site Hamiltonian matrix of the ribbon [see Eq.~\eqref{eq3}]. 

For the uniform case Fig.~\ref{cond} (a) and (b), the conductance plateau sits at $G=2\,e^2/h$, matching the single pair of counter-propagating helical states in the QSHS. This plateau occupies almost the same energy window in both (a) $V=0$ and (b) $V=t$ and show similar overall features.
For the staggered case Fig.~\ref{cond} (c) and (d), the conductance plateau reaches $G=4\,e^2/h$, consistent with the emergence of valley states along with the pseudohelical pair. A narrow, sharp dip also occurs exactly at $E=0$, dropping conductance value down to $G=2\,e^2/h$. This is the transport signature of the small gap identified above for the valley states. For the case (d) V=t, this gap becomes slightly wider. 

We conclude that the charge conductance calculations do not provide a sufficiently sensitive probe to capture the changes in the band dispersion induced by the on-site edge potential for both the uniform and staggered regimes. This motivates us to consider a more direct probe of the spatial and spin-dependent properties of the low-energy states. We therefore turn to RKKY indirect exchange interactions between two magnetic impurities placed within the nanoribbon. Since the RKKY interactions is mediated by the host electrons, it is highly sensitive to their spatial, spin and valley degree of freedoms and provides a more complete probe of the  newly formed valley-polarized edge states.

\subsection{RKKY interaction probe}\label{probeB}
Now we study the signatures of edge states reconstruction via the magnetic exchange interaction (RKKY) by placing two magnetic impurities at the edge of the zigzag nanoribbon.
%-----------------------------------------------------------------
\subsubsection{Theoretical formalism of RKKY interaction}\label{probeB1}
%-----------------------------------------------------------------
In our analysis,  we consider two magnetic impurities placed at $(m_1,n_1)$ and $(m_2,n_2)$ sites of the nanoribbon, following the notations introduced in Fig.~\ref {ribbon}. An indirect exchange interaction is established between magnetic impurities through their coupling with the conduction electrons of the host material \cite{Rudermann,kasuya,Yosida}. The exchange interaction between the impurity magnetic moments and the spins of the conduction electrons is described by the following Hamiltonian\cite{Zare}
\begin{equation}
\mathcal{H}_{\mathrm{int}}=J_c\sum_{j=1,2}\mathbf{S}(\mathbf{R}_j)\cdot\mathbf{I}_j,
\tag{9}
\end{equation}
Here, $J_c$ denotes the coupling constant, $\mathbf{I}_j$ is the magnetic moment associated with the impurity positioned at $\mathbf{R}_j$, spin density of the conduction electrons at position $\mathbf{R}$ is given by $\mathbf{S}(\mathbf{R})=(\hbar/2)\sum_i\delta(\mathbf{r}_i-\mathbf{R})\boldsymbol{\sigma}_i$ , where $\boldsymbol{\sigma}_i$ is the spin density operator acting on the $i$-th conduction electron. 

By applying the second-order perturbation theory, $\mathcal{H}_{\mathrm{int}}$ can be reduced to an effective RKKY interaction given by
\begin{equation}
\mathcal{H}_{\rm RKKY}^{\alpha\beta}
=J_c^2\sum_{k,l}\mathbf{I}_{1k}\chi^{\alpha\beta}_{kl}(\mathbf{r},\mathbf{r}')\mathbf{I}_{2l},
\tag{10}
\end{equation}
Here, $\alpha$ and $\beta$ represent the sublattices hosting the impurities and $\chi_{kl}$ ($k,l=x,y,z$) is the real space spin susceptibility tensor. It can be obtained from the retarded Green's functions $G_{\alpha\beta}^{0}$ as\cite{Zare},
\begin{equation}\label{sus}
\begin{aligned}
\chi_{k,l}^{\alpha\beta}(\mathbf{r},\mathbf{r}')
=-\frac{2}{\pi}\operatorname{Im}
\int_{-\infty}^{\varepsilon_F}d\varepsilon
\operatorname{Tr}\Big[
\sigma^kG_{\alpha\beta}^{0}(\mathbf{r},\mathbf{r}',\varepsilon)\\
\times \sigma^lG_{\beta\alpha}^{0}(\mathbf{r}',\mathbf{r},\varepsilon)\Big],
\end{aligned}
\tag{11}
\end{equation}
where $\varepsilon_F$ is the Fermi energy, and the trace is taken over the spin indices. 

We use the spectral representation of the non-interacting Green's function, to determine the spin susceptibility of the finite-size zigzag nanoribbon\cite{sherafati2012}, 
\begin{equation}
G^{0}_{\alpha\beta}(\mathbf r,\mathbf r',\varepsilon)=\sum_{n}\frac{\psi_n^{\alpha}(\mathbf r)\,\psi_n^{\beta\dagger}(\mathbf r')}{\varepsilon-\varepsilon_n+i\eta},
\tag{12}
\label{greenfunc}
\end{equation}
where the summation runs over the complete set of eigenstates $\psi_n$ and eigenvalues $\varepsilon_n$ obtained by diagonalizing the finite-size Hamiltonian in Eq.~(\ref {eq:Hamiltonian}). Here $\psi_n^{\alpha}(\mathbf r)$ is the two-component spinor of eigenstate $n$ at site $\mathbf r$ on sublattice $\alpha$. Since the Rashba term mixes the spin projections, the eigenstates are not spin polarized and $G^{0}_{\alpha\beta}$ is a $2\times2$ matrix in spin space. Finally, the RKKY Hamiltonian can be written as\cite{Mohsen2}
\begin{equation}
\mathcal{H}_{\mathrm{RKKY}}^{\alpha\beta}
=
\sum_{l,m}
J_{l m}^{\alpha\beta}
\left[
\mathbf{I}_1^l \mathbf{I}_2^m + \mathbf{I}_1^m \mathbf{I}_2^l\right]
+
\vec{J}_{\mathrm{DM}}^{\alpha\beta}
\cdot
\left(\vec{\mathbf{I}}_1 \times \vec{\mathbf{I}}_2\right),
\tag{13}
\end{equation}

These terms can be decomposed into the isotropic Heisenberg interaction $(J_{xx}^{\alpha\beta},J_{yy}^{\alpha\beta})$, Ising interaction $(J_{zz}^{\alpha\beta})$, anisotropic symmetric
interactions $(J_{xy}^{\alpha\beta},J_{yz}^{\alpha\beta},J_{zx}^{\alpha\beta})$, and anisotropic antisymmetric Dzyaloshinskii--Moriya (DM) interactions\cite{Mohsen2} $(J_{\mathrm{DM},x}^{\alpha\beta}, J_{\mathrm{DM},y}^{\alpha\beta}, J_{\mathrm{DM},z}^{\alpha\beta})$. All effective $J$'s are proportional to the static spin susceptibility tensor of the system \cite{schulz}, with $J_{mm}^{\alpha\beta}\propto \chi_{mm}^{\alpha\beta}$, $J_{lm}^{\alpha\beta} \propto (\chi_{lm}^{\alpha\beta}+\chi_{ml}^{\alpha\beta})/2$ for $l\neq m$, and $J_{\mathrm{DM}}^{k,\alpha\beta} \propto \sum_{l,m}\epsilon_{klm}\chi_{lm}^{\alpha\beta}/2$. The explicit form of the spin susceptibility tensor used in our numerical calculations, is given in the \hyperref[appendix]{Appendix}.

%------------------------------------------------------
\subsubsection{Results and Discussion}\label{probeB2}
%------------------------------------------------------
\begin{figure*}[tb]
    \centering
     \includegraphics[width=0.46\linewidth]{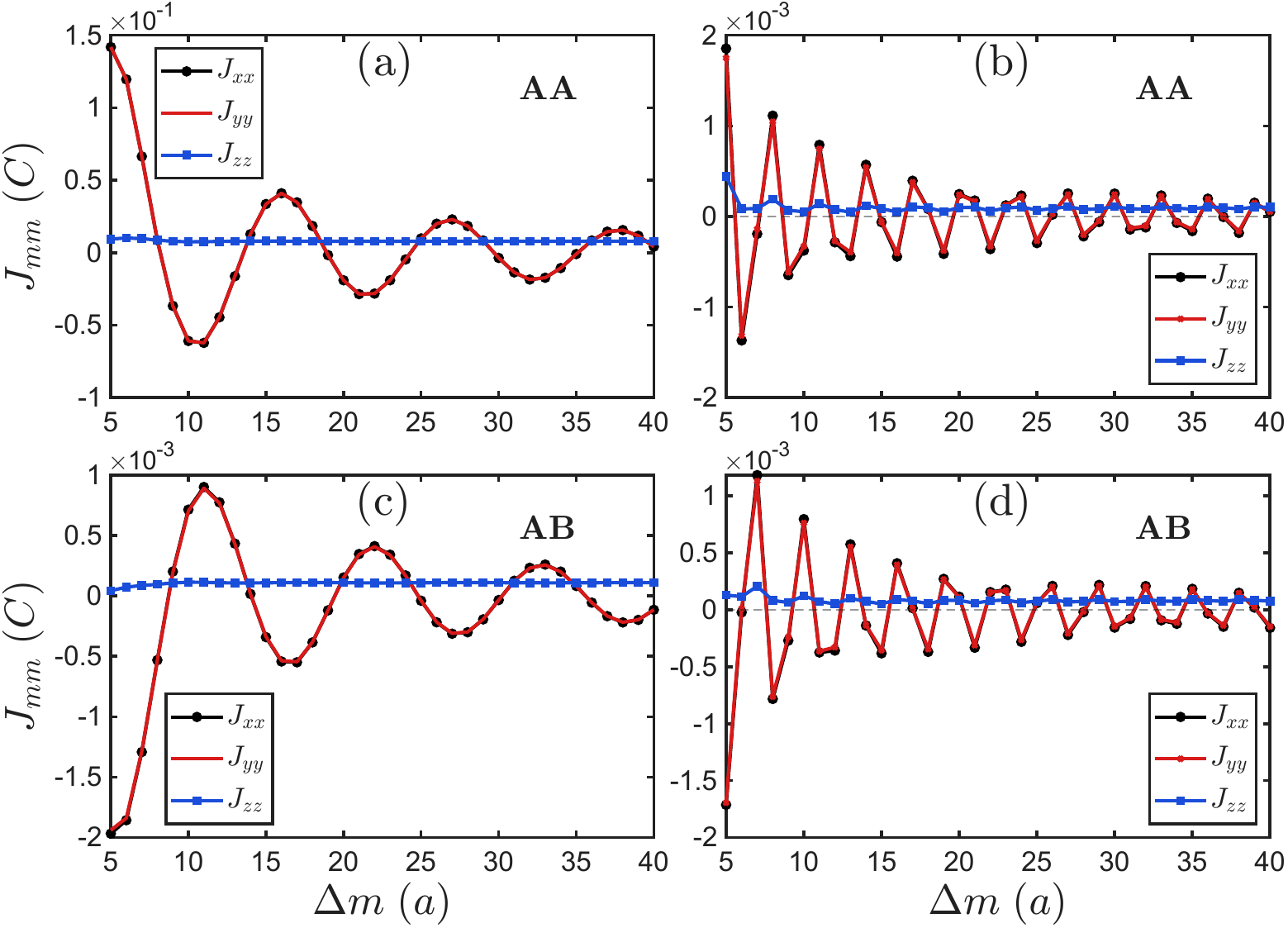}
     \hspace{4mm}
     \includegraphics[width=0.46\linewidth]{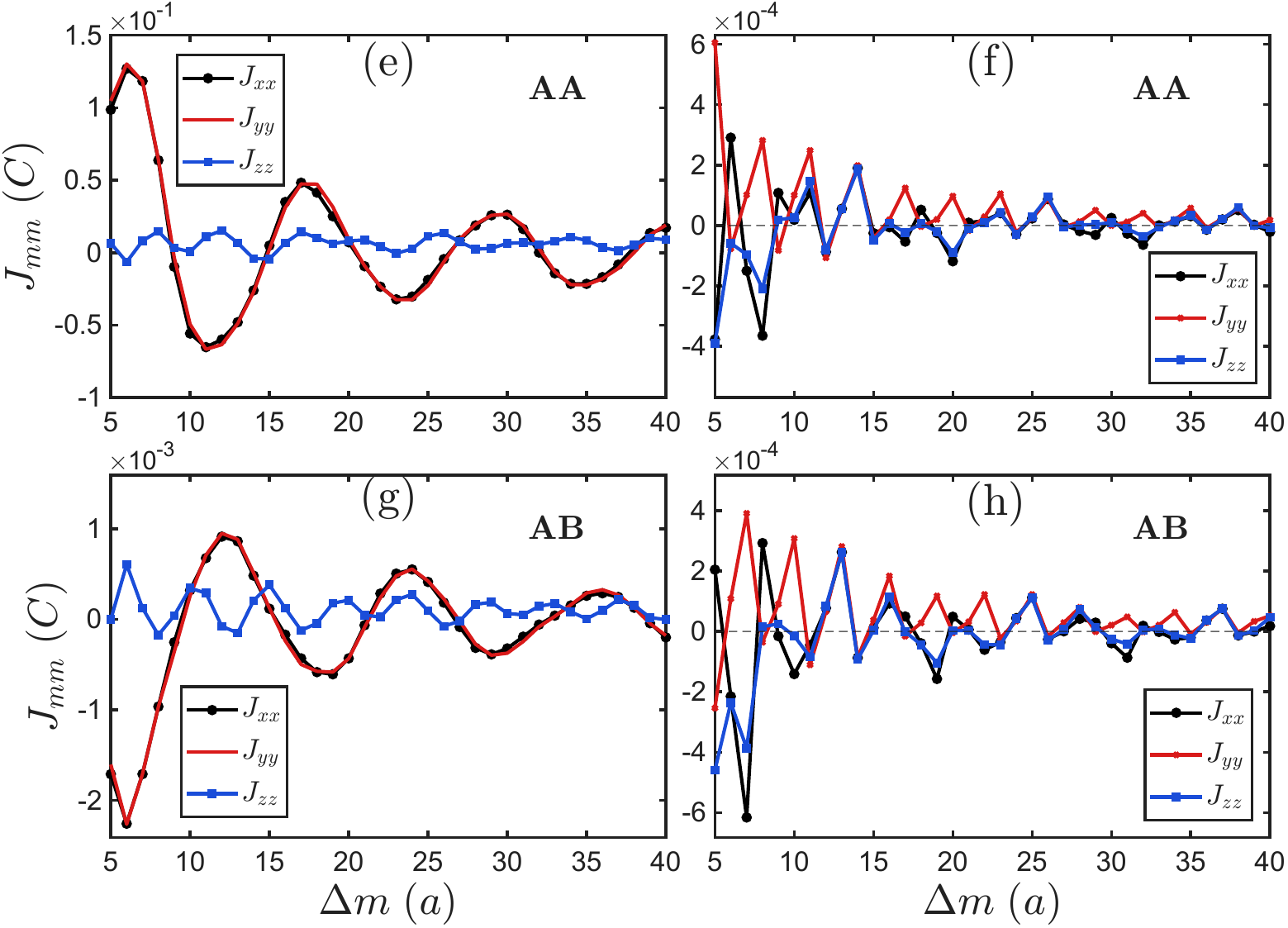}
    
   \caption{(Color online) The Heisenberg interaction terms scaled by $J_c^2$ for $N_y=60$ and $L=80$  between two magnetic impurities are plotted as a function of the distance between them (in units of the lattice parameter $a$). Left and right columns depict the uniform and staggered SOC cases, respectively. In all the top panels both the impurities occupy same sublattice sites (AA), and for the bottom panels they occupy different sublattice sites (AB). The edge potential is fixed at V=0 in panels (a)(c),(e), and (g) and V = t in panels (b),(d),(f),and (h).  }
    \label{fig4}
\end{figure*}

\begin{figure*}[t]
    \centering
     \includegraphics[width=0.46\linewidth]{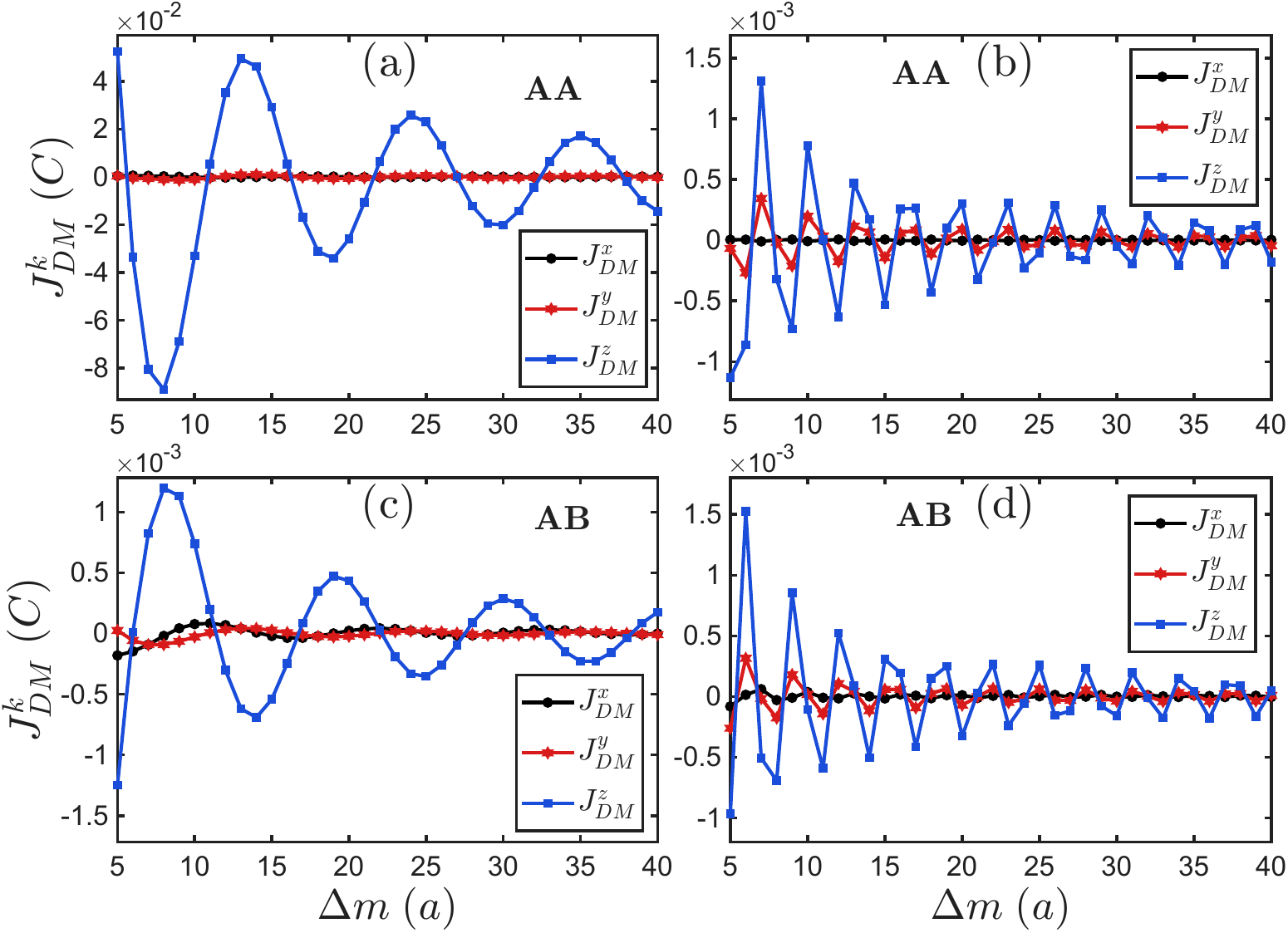}
     \hspace{4mm}
     \includegraphics[width=0.46\linewidth]{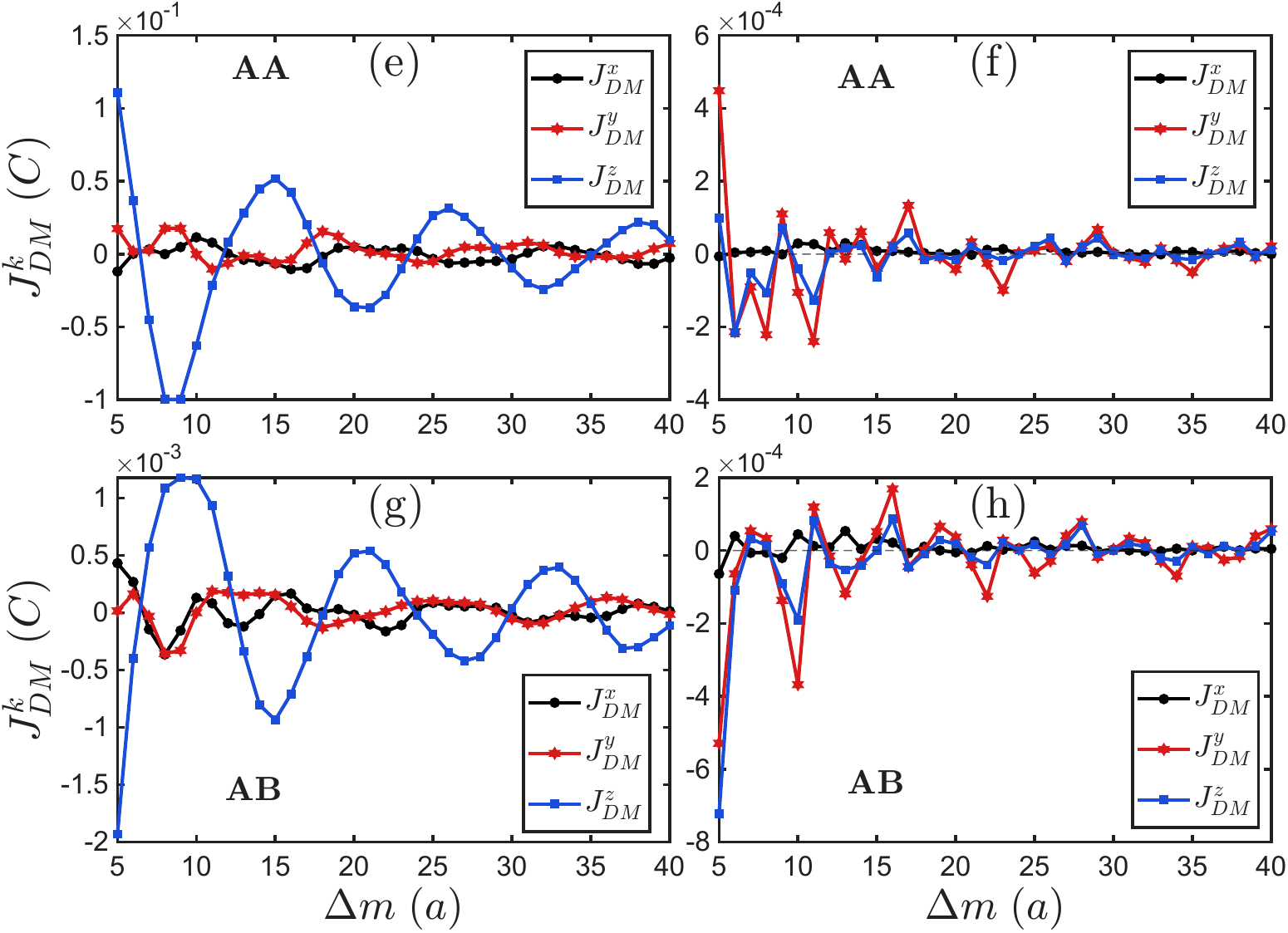}
    \caption{(Color online) The DM interaction terms scaled by $J_c^2$ for $N_y=60$ and $L=80$  between two magnetic impurities are plotted as a function of the distance between them (in units of the lattice parameter $a$). Left and right columns depict the uniform and staggered SOC cases, respectively. In all the top panels both the impurities occupy same sublattice sites (AA), and for the bottom panels they occupy different sublattice sites (AB). The edge potential is fixed at V=0 in panels (a)(c),(e), and (g) and V = t in panels (b),(d),(f),and (h).  }
    \label{fig5}
\end{figure*}

In this section, we present our numerical results for various RKKY exchange interaction terms and their dependence on the distance between the two magnetic impurities for the uniform and staggered SOC cases. We consider two different situations when both the impurities are located at the same sublattices (AA) of the nanoribbon or they are situated at different sublattices (AB). We fix one of the impurities at the position $(10,1)$ and the location of the second impurity is at $(m,1)$ for AA case. For the AB case, the second impurity is located along the line $n=2$ at $(m,2)$. We use Eq.~(\ref{sus}) for numerically evaluating different components of the sub-lattice resolved RKKY exchange interaction. We vary $\Delta m \; (= 10 \;\mathord{-}\;m)$ from 5 to 40 (in units of the lattice parameter a). We keep the chemical potential at $\mu=0$ as our aim is to only extract the edge state's signature, and not to mix up with the bulk states. In order to probe the edge potential induced edge state reconstruction we consider both the cases i.e., with and without edge potential.  The size of the ribbon is considered as: length $L=80$ and width $N_y=60$. Increasing the ribbon length $L$ or width $N_y$ further does not change the qualitative features of our results.
% HEISENBERG
\begin{figure*}[tb]
    \centering
     \includegraphics[width=0.46\linewidth]{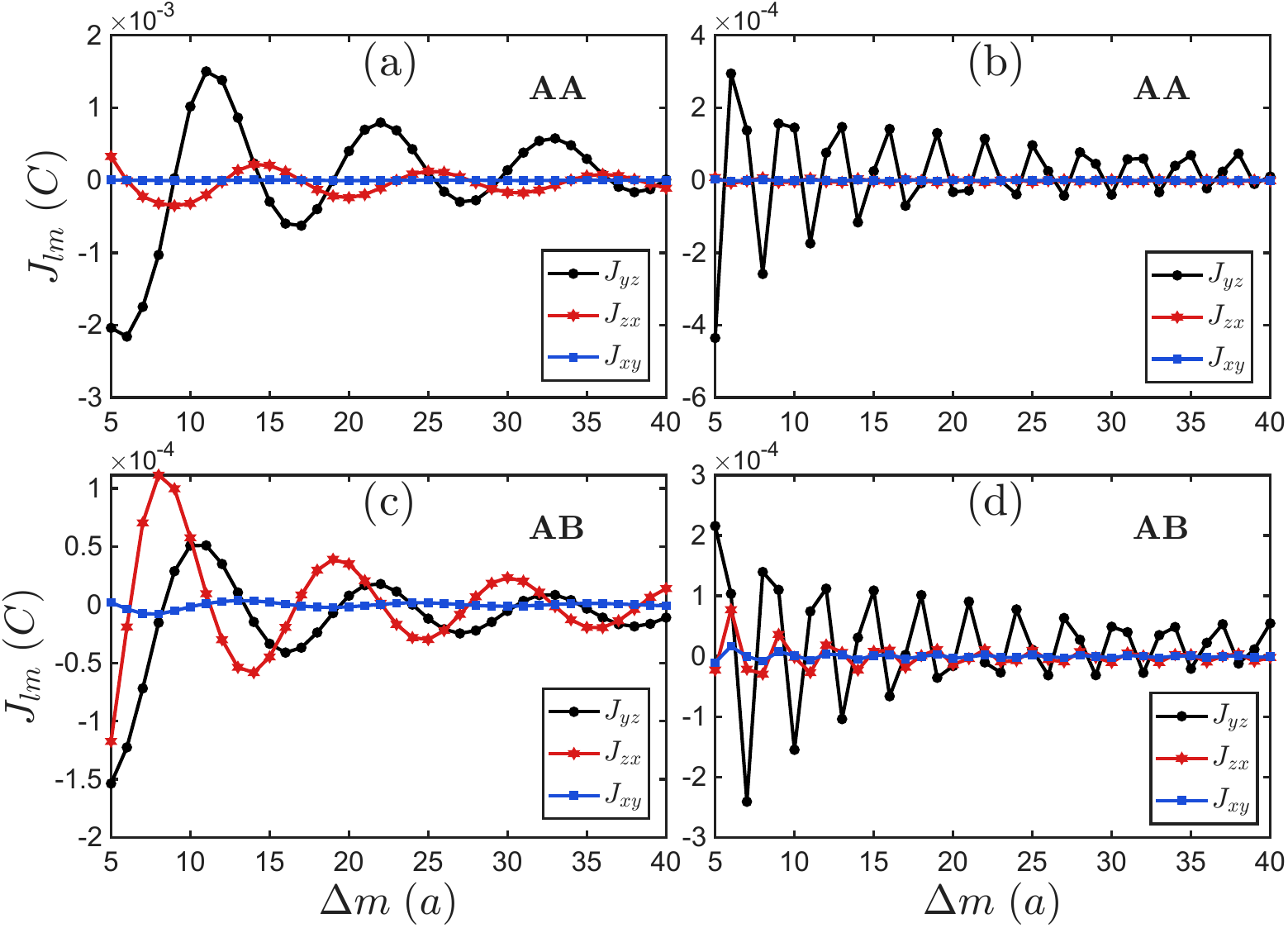}
     \hspace{4mm}
     \includegraphics[width=0.46\linewidth]{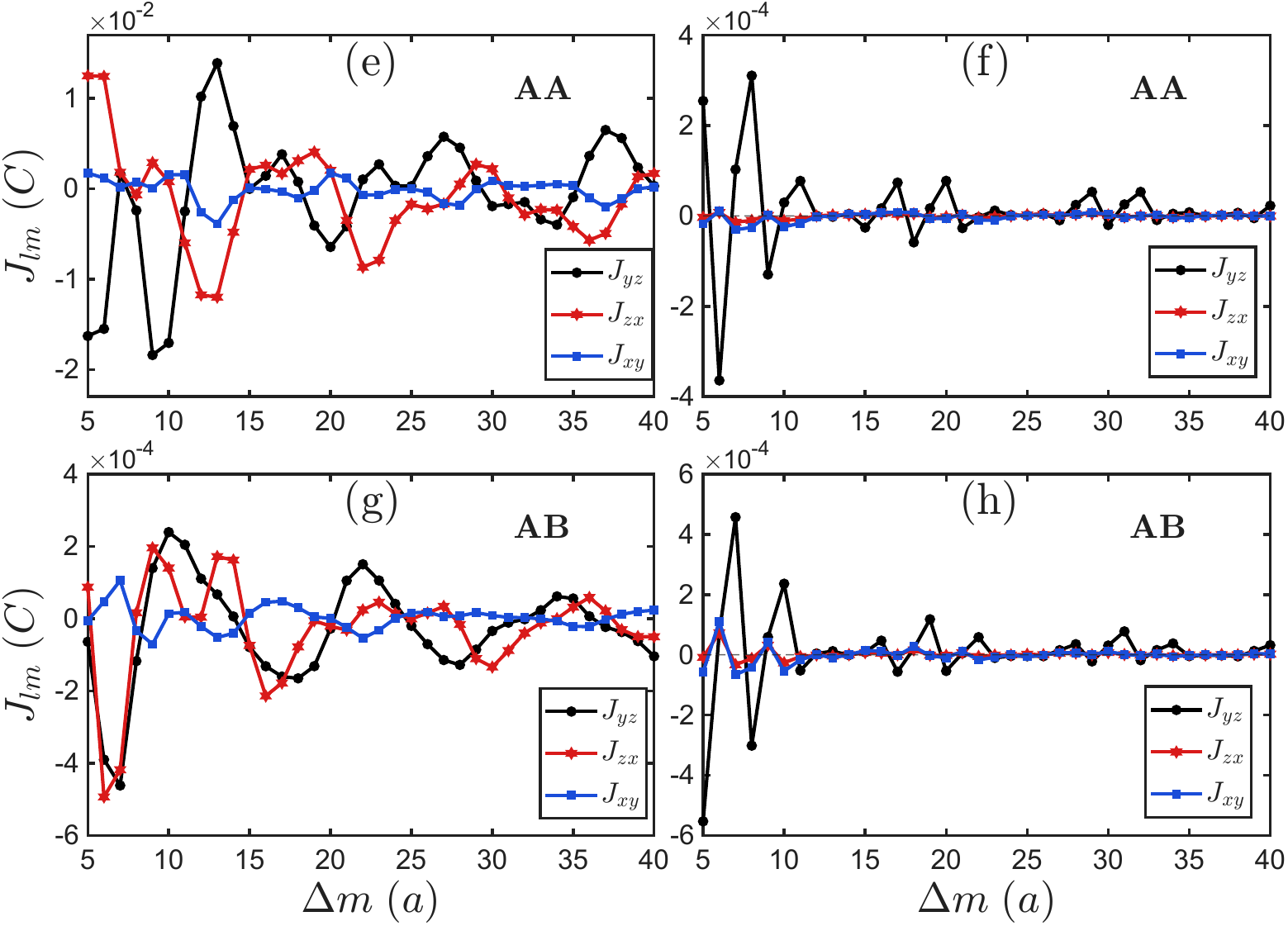}
    \caption{(Color online) The symmetric interaction terms scaled by $J_c^2$  for $N_y=60$ and $L=80$  between two magnetic impurities are plotted as a function of the distance between them (in units of the lattice parameter $a$). Left and right columns depict the uniform and staggered SOC cases, respectively. In all the top panels both the impurities occupy same sublattice sites (AA), and for the bottom panels they occupy different sublattice sites (AB). The edge potential is fixed at V=0 in panels (a)(c),(e), and (g) and V = t in panels (b),(d),(f),and (h).  }
    \label{fig6}
\end{figure*}

In Figure  \ref{fig4}, we plot the diagonal components of the RKKY exchange tensor, $J_{mm}\propto\chi_{mm}$ ($m=x,y,z$). The left and right columns depict the uniform and staggered spin-orbit coupling regimes, respectively, as a function of impurity separation $\Delta m$. Each column again contains left and right sub-columns, corresponding to $V=0$ and $V=t$, respectively. In absence of edge potential, $J_{xx}=J_{yy}$ varyies smoothly with $\Delta m$ while $J_{zz}$ is strongly suppressed in comparison the the other components. [see Figs.~\ref{fig4} (a),(c) and ~\ref{fig4} (e),(g)]. This is expected for ordinary helical edge states whose spin is locked to momentum\cite{Verma,Gao}. The {\rm AA} and {\rm AB} curves differ in overall phase and magnitude, with $J_{xx}=J_{yy}$ reversing sign character between the two sublattice configurations and AB being smaller in magnitude by two orders. Here, both AA and AB curves oscillate between ferromagnetic (positive) and anti-ferromagnetic (negative) character with distance, but oppositely, so that the two channels remain phase-reversed relative to one another. This is consistent with the general result for the undoped case that the RKKY interaction on a bipartite lattice is ferromagnetic for same-sublattice pairs and antiferromagnetic for opposite-sublattice pairs as a consequence of particle-hole symmetry at the Dirac point \cite{Saremi,Annica,SherafatiDoped}. At $V=0$ both the regimes have $J_{xx}=J_{yy}\gg J_{zz}$ suggesting that the effective spin-quantization axis along $z$ remains approximately preserved irrespective of the SOC staggering pattern in absence of edge potential, though $J_{zz}$ already develops a weak oscillatory component in the staggered case that is absent in the uniform case. 

Applying a finite edge potential $V=t$ shifts the edge states toward the valley gap [Figs.~\ref{band} (b) and (d)]. The AA/AB phase reversal is absent now, in both the uniform [Figs.~\ref{fig4} (b) and (d)] and staggered regimes [Figs.~\ref{fig4} (f) and (h)]. All the $V=t$ cases additionally show a higher oscillation frequency than the corresponding $V=0$ curves, consistent with the emergence of additional oscillatory contributions due to edge potential. The faster oscillation is the result of momentum-space reconstruction of the edge states' Green function entering into the spin susceptibility in Eq. ~(\ref{sus}). In presence of the edge potential,  for the uniform SOC case, $J_{xx}$ and $J_{yy}$ are degenerate and $J_{zz}$ is vanishingly small compared to the other two components [Fig.~\ref{fig4} (b) and (d)]. In the staggered case, by contrast, applying edge potential lifts the $J_{xx}=J_{yy}$ degeneracy and  $J_{zz}$ becomes comparable to the in-plane components [Fig.~\ref{fig4} (f) and (h)] when both the intrinsic sublattice asymmetry and edge potential act together. 
%DM TERMS

Figure~\ref{fig5} shows the three components of the DM exchange interaction, $J_{DM}^k$ ($k=x,y,z)$, for the uniform (left column) and staggered (right column) SOC regimes. At $V=0$ in the uniform case [Figs.~\ref{fig5} (a) and (c)], $J^x_{DM}$ and $J^y_{DM}$ are strongly suppressed relative to $J^z_{DM}$ for AA sites, leaving $J^z_{DM}$ as the dominant component. This reflects the fact that intrinsic (Kane-Mele) SOC conserves $s_z$\cite{Kane}. Placing the impurities on opposite sublattices (AB) weakens this symmetry, allowing the in-plane components to grow relative to $J^z_{DM}$. The scale for AB case is again an order of magnitude smaller than the AA case. The small residual in-plane components likely arise from the Rashba spin-orbit interaction, which is known to favor an in-plane DM vector \cite{Imamura}.  At $V=t$ [Figs.~\ref{fig5} (b) and (d)], $J^y_{DM}$ grows to become comparable to $J^z_{DM}$ in both AA and AB case, while $J^x_{DM}$ remains suppressed. 
Since the edge potential does not itself couple to spin, this suggests that whatever approximate symmetry keeps $J^x_{DM}$ small is largely insensitive to edge potential, unlike the symmetry corresponding to $J^y_{DM}$, which the edge potential clearly disrupts.

In the staggered case , by contrast, $J_{\mathrm{DM}}^{x}$ is comparable to $J_{\mathrm{DM}}^{y}$ already at $V=0$ [Figs.~\ref{fig5} (e) and (g)] while $J_{\mathrm{DM}}^{z}$ still remains dominant. Also, the amplitude of DM terms in presence of staggered SOC, is roughly an order of magnitude larger than that in the uniform case. At $V=t$  [Figs.~\ref{fig5} (f) and (h)], all three components decrease in magnitude and become comparable indicating that no single axis remains preferred once both staggering and edge potential act together. Similar to the diagonal terms all the $V=t$ curves show higher oscillation frequency than the corresponding $V=0$ curves.

%SYMMETRIC TERMS
Figure~\ref{fig6}  show the off-diagonal symmetric components $J_{xy}$, $J_{yz}$, $J_{zx}$ for uniform (left column) and staggered (right column) SOC cases. In the uniform case, at $V=0$  [Figs.~\ref{fig6} (a) and (c)], $J_{yz}$ dominates while $J_{zx}$ is smaller but active, and $J_{xy}$ stays close to zero. For AB, $J_{zx}$ becomes the dominant term, with a comparable $J_{yz}$ and negligible $J_{xy}$, additionally the magnitude of AB is an order less than AA. At $V=t$  [Figs.~\ref{fig6} (b) and (d)], $J_{yz}$ remains dominant with higher oscillation frequency while $J_{zx}$ and $J_{xy}$ are strongly suppressed for both AA/AB cases.

In the staggered case at $V=0$   [Figs.~\ref{fig6} (e) and (g)], all three components become active and comparable in size, with no clear single dominant term, and the  scale for AA is roughly an order of magnitude larger than in the uniform case. At $V=t$,  staggered SOC [Figs.~\ref{fig6} (f) and (h)] shows another distinct feature where instead of the smooth, monotonically decaying oscillations seen in the uniform case with edge potetntial, $J_{yz}$ shows a beating pattern for both AA/AB cases.
%-----------------------------------------------
\section{Conclusion}\label{conc}
To conclude, we studied the effect of edge potential on the band dispersion of proximitized graphene in the presence of sublattice asymmetry and staggered on-site potential. The band dispersion is sensitive to the applied edge potential and drives the helical and pseudo-helical edge states towards purely intra-valley states, exhibiting both spin and valley polarization. We found that the two-terminal charge conductance fails to capture this reconstruction, since it depends only on the number of transmitted channels, which remains unchanged. We therefore probed this shift using RKKY interactions, which decay with a significantly higher oscillation frequency in the presence of edge potential. The RKKY interaction comprises several contributions, namely Heisenberg, Ising, symmetric anisotropic, and Dzyaloshinskii-Moriya (DM) terms. The dominant components among various contributions are governed by the underlying symmetries of the system. We also find that the RKKY interactions are stronger when the two impurities sit on the same sublattice sites, while a sign reversal between AA and AB configurations occurs for the zero edge potential case, in the Heisenberg and DM terms. The amplitude of DM terms is enhanced in the presence of staggered SOC and non-zero in-plane components emerge. These results establish RKKY interactions as a spin and sublattice resolved probe of proximity induced effects in graphene nanoribbons, of potential relevance to future spintronic and valleytronic device applications.
%-----------------------------------------------

\section*{Acknowledgement}
SK Firoz Islam acknowledges financial support from the project ANRF/ECRG/2024/005166/PMS.

\section*{Data Availability}
The data that supports the findings of this article are not publicly available. The data are available from the authors upon reasonable request.
\vspace{3mm}

\section*{appendix}\label{appendix}
Here we derive the explicit form of the spin susceptibility, Eq.~\eqref{sus}, used in our numerical calculations. In Ref.~[\onlinecite{Zare}],  the Rashba term is absent and $S_z$ is conserved, so every eigenstate is either spin-up or spin-down. In the present case, Rashba coupling mixes the two spin projections and the eigenstates  are no longer spin polarized. As a result the Green's function does not separate into spin-up and spin-down blocks, and all nine components $\chi_{kl}$ ($k,l=x,y,z$) must be retained.

We write the eigenstates, obtained by exact diagonalization of the finite ribbon, as $\psi_n(\mathbf r)=\big(\psi_n^{\uparrow}(\mathbf r),\psi_n^{\downarrow}(\mathbf r)\big)^{T}$, with energies $\varepsilon_n$. Inserting the spectral representation of the retarded Green's function,
Eq.~(\ref{greenfunc}),  in Eq.~(\ref{sus}) we obtain,
\begin{multline}
\operatorname{Tr}\!\big[\sigma^{k}\psi_n(\mathbf r)\psi_n^{\dagger}(\mathbf r')\,\sigma^{l}\psi_m(\mathbf r')\psi_m^{\dagger}(\mathbf r)\big]\\
=\big[\psi_m^{\dagger}(\mathbf r)\sigma^{k}\psi_n(\mathbf r)\big]\big[\psi_n^{\dagger}(\mathbf r')\sigma^{l}\psi_m(\mathbf r')\big]\equiv M^{kl}_{nm}(\mathbf r,\mathbf r'),
\label{eq:M}
\tag{A1}
\end{multline}
so that
\begin{multline}
\chi_{kl}(\mathbf r,\mathbf r')=-\frac{2}{\pi}\operatorname{Im}\sum_{n,m}M^{kl}_{nm}(\mathbf r,\mathbf r')\\
\times\int_{-\infty}^{\varepsilon_F}\frac{d\varepsilon}{(\varepsilon-\varepsilon_n+i\eta)(\varepsilon-\varepsilon_m+i\eta)}.
\label{eq:chiM}
\tag{A2}
\end{multline}

 At zero temperature, with $f(\varepsilon)=\Theta(\varepsilon_F-\varepsilon)$, and using $\lim_{\eta\to0^{+}}(x+i\eta)^{-1}=\mathcal P(1/x)-i\pi\delta(x)$, after some algebra, we obtain 
\begin{multline}
\operatorname{Im}\int_{-\infty}^{\varepsilon_F}\frac{d\varepsilon}{(\varepsilon-\varepsilon_n+i\eta)(\varepsilon-\varepsilon_m+i\eta)}\\
=-\pi\,\frac{f(\varepsilon_n)-f(\varepsilon_m)}{\varepsilon_n-\varepsilon_m}.
\label{eq:Iint}
\tag{A3}
\end{multline}
Here $M^{kl}_{mn}=\big(M^{kl}_{nm}\big)^{*}$ and the energy integral is symmetric under $n\leftrightarrow m$, so that the imaginary part of $M^{kl}_{nm}$ cancels, leaving

\begin{multline}
\chi_{kl}(\mathbf r,\mathbf r')=2\sum_{n\neq m}\frac{f(\varepsilon_n)-f(\varepsilon_m)}{\varepsilon_n-\varepsilon_m}\\
\times\operatorname{Re}\Big\{\big[\psi_m^{\dagger}(\mathbf r)\,\sigma^{k}\,\psi_n(\mathbf r)\big]\big[\psi_n^{\dagger}(\mathbf r')\,\sigma^{l}\,\psi_m(\mathbf r')\big]\Big\}.
\label{eq:chi_final}
\tag{A4}
\end{multline}

  %\begin{acknowledgments}
 %\end{acknowledgments}
 \bibliography{References}
 \end{document}